%% file: loop_gravity.tex
\documentclass[twocolumn,english,nofootinbib]{revtex4-1}
\usepackage[T1]{fontenc}
\usepackage[latin9]{inputenc}
\usepackage[a4paper]{geometry}
\usepackage{amsmath}
\usepackage{amssymb}
\usepackage{cancel}
\usepackage{stmaryrd}
\usepackage[dvipsnames]{xcolor}
\usepackage[normalem]{ulem}
\usepackage{tikz}

\makeatletter
\usepackage{hyperref}
\usepackage{slashed}
\usepackage{shuffle}

\makeatother

\usepackage{babel}
\begin{document}
\title{One-loop $N$-point correlators in pure gravity}
\author{Humberto Gomez$^{a,b}$}
\author{Renann Lipinski Jusinskas$^{b}$}
\author{Cristhiam Lopez-Arcos$^{c,d}$}
\author{Alexander Quintero V\'elez$^{c}$}
\affiliation{$^{a}$ Facultad de Ciencias Basicas, Universidad Santiago de Cali,~\\
Calle 5 $N^\circ$ 62-00 Barrio Pampalinda, Cali, Valle, Colombia}
\affiliation{$^{b}$ Institute of Physics of the Czech Academy of Sciences \& CEICO
~\\
Na Slovance 2, 18221 Prague, Czech Republic}
\affiliation{$^{c}$ Escuela de Matem\'{a}ticas, Universidad Nacional de Colombia, \\ Sede Medell\'{i}n, Carrera 65 $\#$ 59A--110, Medell\'{i}n, Colombia}
\affiliation {$^{d}$ Universidad EIA, C.P. 055428,  Envigado, Colombia}

\begin{abstract}
In this work we propose a simple algebraic recursion for the complete one-loop integrands of $N$-graviton correlators. This formula automatically yields the correct symmetry factors of individual diagrams, taking into account both the graviton and the ghost loop, and seamlessly controlling the related combinatorics.
\end{abstract}
\maketitle

\section{Introduction}

Achieving a satisfactory quantum description of gravity has long been a major aspiration in theoretical physics. As a quantum field theory, earlier research has introduced seminal works (e.g. \cite{Feynman:1996kb,Weinberg:1964kqu,Weinberg:1964ew,Weinberg:1965rz,Weinberg:1965nx,DeWitt:1967yk,DeWitt:1967ub,DeWitt:1967uc}), but we were soon confronted with an inconvenient observation: gravity is not renormalizable \cite{tHooft:1974toh}. From an effective field theory (EFT) perspective, this is not an obstacle but rather a source of valuable lessons using perturbation theory (see \cite{Donoghue:2022eay} and references therein). 

In ordinary gauge theories, perturbative computations using a traditional diagrammatic approach are incumbered by gauge redundancies. This is even more pronounced in gravity, which in addition contains an infinite number of field theory vertices and non-trivial field redefinitions (e.g. \cite{Ananth:2007zy,Knorr:2023usb}). While these redundancies appear to be unavoidable in a Lagrangian formulation, modern scattering amplitudes techniques have long abandoned them. The so-called on-shell methods (see e.g. \cite{Travaglini:2022uwo,Brandhuber:2022qbk,Bern:2022wqg}) have revolutionized the area, enabling the systematic computation of S-matrices based on just a  few ingredients (particle content, global symmetries, factorization, unitarity). A paradigmatic example is the double copy \cite{Bern:2008qj,Bern:2010ue}.

Beyond on-shell results, there is increased interest on graviton correlators in different contexts: renormalization group analysis, form factors, curved backgrounds (including AdS/CFT), and gravitational EFTs, to name a few. Since the role of gravitons is firmly established in modern theoretical physics, going off the mass-shell is an important step. This is especially relevant in the exciting new era of gravitational waves and black hole observations. An off-shell formulation would naturally lead to a sounder description of the quanta of gravity, whether or not they can be detected \cite{Rothman:2006fp,Dyson:2013hbl}. At the same time, it could further expose the shortcomings of a particle/field description in lieu of a more overarching model such as string theory \cite{Weinberg:2021exr}.

Current off-shell methods are few and limited. The string-inspired world-line formalism \cite{Schubert:2001he}, for instance, is presently well-suited to compute only the irreducible part of one-loop correlators \cite{Bastianelli:2013tsa,Ahmadiniaz:2012xp,Ahmadiniaz:2020jgo,Ahmadiniaz:2023vrk}. In spite of ongoing interest (e.g. \cite{Bastianelli:2022pqq,Brandt:2022und}), state-of-the-art loop computations with off-shell gravitons are still inneficiently based on diagram expansions, as in \cite{Goroff:1985sz,Goroff:1985th}. Some time ago, we established a recursive technique to compute one-loop correlators in colored theories \cite{Gomez:2022dzk} via an extension of the perturbiner method \cite{Rosly:1996vr,Rosly:1997ap}. The underlying idea is to generate scattering trees from field equations. This has been long known at tree level \cite{Boulware:1968zz} (also \cite{Berends:1987me,Bardeen:1995gk,Cangemi:1996rx}), and the perturbiner streamlined the procedure (see \cite{Mafra:2015gia,Lee:2015upy,Mafra:2015vca,Mafra:2016ltu,Mafra:2016mcc,Mizera:2018jbh,Garozzo:2018uzj,Lopez-Arcos:2019hvg,Gomez:2020vat,Guillen:2021mwp,Gomez:2021shh,Ben-Shahar:2021doh,Cho:2021nim,Escudero:2022zdz,Lee:2022aiu,Cho:2022faq,Cho:2023kux,Damgaard:2024fqj,Chen:2023bji,Tao:2023yxy} for a series of recent developments and applications). The quantum (off-shell) step of \cite{Gomez:2022dzk} involved a controlled sewing procedure to recursively generate one-loop integrands. These results relied heavily on the color ordering (like in bi-adjoint scalar and Yang-Mills theories), and could not be applied to colorless fields.

In this work we finally report a breakthrough in the recursive construction of one-loop $N$-graviton correlators. The perturbiner offers a natural solution to the problem of defining a recursion encompassing the infinite number of graviton vertices \cite{Gomez:2021shh}. Our proposal is based on an elegant solution to the expected overcounting of Feynman diagrams generated by the sewing procedure, ultimately taming their combinatorial nature. We show that our formula automatically generates the fully off-shell integrands, with both graviton and ghost loops. As a side result, we also propose a formula for one-loop amplitudes, automatically removing tadpoles and external-leg bubbles.


\section{Equations of motion}

The Einstein-Hilbert action in $d$ dimensions is
\begin{equation}
S_{\text{EH}}=\frac{1}{2\kappa}\int d^{d}x\sqrt{-g}R,\label{eq:EH-action}
\end{equation}
where $\kappa$ is the gravitational constant, $g$ is the determinat
of the inverse metric $g_{\mu\nu}$, and $R=g^{\mu\nu}R_{\mu\nu}$
is the scalar curvature. The Ricci tensor is expressed as $R_{\mu\nu}=R_{\mu\rho\nu}^{\rho}$,
with Riemann tensor given by
\begin{equation}
R_{\mu\rho\nu}^{\sigma} = \partial_{\rho}\Gamma_{\mu\nu}^{\sigma}-\partial_{\nu}\Gamma_{\rho\mu}^{\sigma}+\Gamma_{\rho\gamma}^{\sigma}\Gamma_{\mu\nu}^{\gamma}-\Gamma_{\nu\gamma}^{\sigma}\Gamma_{\mu\rho}^{\gamma},
\end{equation}
and $\Gamma^{\rho}_{\mu\nu} = g^{\rho \sigma}(\partial_{\mu}g_{\nu\sigma}+\partial_{\nu}g_{\mu\sigma}-\partial_{\sigma}g_{\mu\nu})/2$.

Regardless of the natural geometrical interpretation, the action \eqref{eq:EH-action}
and the metric $g_{\mu\nu}$ often make the analysis of
graviton scattering overly involved. We will work instead with the metric density \cite{Landau:1975pou}
\begin{equation}
\mathfrak{g}^{\mu\nu}=\sqrt{-g}g^{\mu\nu},
\end{equation}
with inverse $\mathfrak{g}_{\mu\nu}$. Under diffeomorphisms ($\delta x^{\mu}=\tilde{c}^{\mu}$), it transforms as $\delta \mathfrak{g}^{\mu \nu}=\mathfrak{g}^{\mu \rho} \partial_\rho \tilde{c}^{\nu}+\mathfrak{g}^{\nu \rho} \partial_\rho \tilde{c}^{\mu}-\partial_{\rho} (\mathfrak{g}^{\mu \nu} \tilde{c}^{\rho})$.

Towards gauge-fixing, we introduce the Faddeev--Popov ghost $c^\mu$, with antighost $b_\mu$. The usual harmonic gauge reads $g^{\nu \rho}\Gamma_{\nu \rho}^{\mu}=\partial_{\nu}\mathfrak{g}^{\mu\nu}=0$, and the gauge-fixed action is given by
\begin{multline}
S=\frac{1}{8\kappa}\int d^{d}x\, \big \{ \mathfrak{g}^{\mu\nu}\partial_{\mu}\mathfrak{g}^{\rho\sigma}\partial_{\nu}\mathfrak{g}_{\rho\sigma}-2\mathfrak{g}^{\mu\nu}\partial_{\mu}\mathfrak{g}^{\rho\sigma}\partial_{\rho}\mathfrak{g}_{\nu\sigma}\\
-\frac{1}{(2-d)}\mathfrak{g}^{\mu\nu}\mathfrak{g}^{\rho\sigma}\partial_{\mu}\mathfrak{g}_{\rho\sigma}\mathfrak{g}^{\gamma\lambda}\partial_{\nu}\mathfrak{g}_{\gamma\lambda} -2\eta_{\mu\nu}\partial_{\rho}\mathfrak{g}^{\mu\rho}\partial_{\sigma}\mathfrak{g}^{\nu\sigma}\\ \vphantom{\frac{1}{(2-d)}}
+4[\mathfrak{g}^{\mu\rho}\partial_{\rho}c^{\nu}+\mathfrak{g}^{\nu\rho}\partial_{\rho}c^{\mu}-\partial_{\rho}(\mathfrak{g}^{\mu\nu}c^{\rho})]\partial_{\mu}b_{\nu}\big \},\label{eq:fixed-action}
\end{multline}
where $\eta_{\mu\nu}$ is the flat space metric.

The respective equations of motion are straightforward to derive. For the metric density we obtain
\begin{multline}
\partial_{\rho}(\mathfrak{g}^{\rho\sigma}\partial_{\sigma}\mathfrak{g}_{\mu\nu})=\eta_{\mu\sigma}\partial_{\nu}\partial_{\rho}\mathfrak{g}^{\rho\sigma}-\partial_{\rho}(\mathfrak{g}_{\mu\sigma}\partial_{\nu}\mathfrak{g}^{\rho\sigma})\\
+\frac{1}{4}\partial_{\mu}\mathfrak{g}^{\rho\sigma}\partial_{\nu}\mathfrak{g}_{\rho\sigma}+\frac{1}{2}\mathfrak{g}^{\rho\sigma}\mathfrak{g}^{\gamma\lambda}\partial_{\gamma}\mathfrak{g}_{\mu\rho}(\partial_{\lambda}\mathfrak{g}_{\nu\sigma}-\partial_{\sigma}\mathfrak{g}_{\nu\lambda})\\
-\frac{1}{4}\frac{1}{(2-d)}[2\mathfrak{g}_{\mu\nu}\partial_{\rho}(\mathfrak{g}^{\rho\sigma}\mathfrak{g}^{\gamma\lambda}\partial_{\sigma}\mathfrak{g}_{\gamma\lambda})+\mathfrak{g}^{\rho\sigma}\partial_{\mu}\mathfrak{g}_{\rho\sigma}\mathfrak{g}^{\gamma\lambda}\partial_{\nu}\mathfrak{g}_{\gamma\lambda}]\\
+\frac{1}{2}\partial_{\mu}c^{\rho}\partial_{\nu}b_{\rho}+\frac{1}{2}\partial_{\mu}(c^{\rho}\partial_{\rho}b_{\nu})+(\mu\leftrightarrow\nu). \label{eq:eom-g}
\end{multline}
while the equations of motion for the ghosts are
\begin{eqnarray}
\partial_{\nu}(\mathfrak{g}^{\nu\rho}\partial_{\rho}c^{\mu})-\partial_{\rho}(\partial_{\nu}\mathfrak{g}^{\mu\nu}c^{\rho}) & = & 0,\label{eq:eom-c}\\
\partial_{\nu}(\mathfrak{g}^{\nu\rho}\partial_{\rho}b_{\mu})+\partial_{\nu}\mathfrak{g}^{\nu\rho}\partial_{\mu}b_{\rho} & = & 0.\label{eq:eom-b}
\end{eqnarray}

We can use them to conveniently extract scattering trees through the gravitational perturbiner  \cite{Gomez:2021shh}. The so-called multiparticle expansions of the fields $\mathfrak{g}_{\mu\nu}$,
$\mathfrak{g}^{\mu\nu}$, $b_{\mu}$, and $c^{\mu}$ can be cast as
\begin{eqnarray}
\mathfrak{g}_{\mu\nu}(x) & = & \eta_{\mu\nu}+\sum_{P}H_{P\mu\nu}e^{ik_{P}\cdot x},\\
\mathfrak{g}^{\mu\nu}(x) & = & \eta^{\mu\nu}-\sum_{P}I_{P}^{\mu\nu}e^{ik_{P}\cdot x},\\
c^{\mu}(x) & = & \sum_{P}C_{P}^{\mu}e^{ik_{P}\cdot x},\\
b_{\mu}(x) & = & \sum_{P}B_{P\mu}e^{ik_{P}\cdot x},
\end{eqnarray}
These expansions are given in terms of plane waves, with single-particle states as building blocks, and the metric density is expanded around flat space. $P=p_1 p_2 \ldots p_n$ denotes an ordered
word composed of $n$ single-particle labels (letters) $p_i$, with $p_i <p_{i+1}$. The sum over words, $\sum_{P}$,
goes from single-particle states (one-letter words) to any multiparticle
composition. The multiparticle momenta $k_{P}$ are defined as $k_{P}=k_{p_{1}}+...+k_{p_{n}}$. The multiparticle
currents ($H_{P\mu\nu}$, $I_{P}^{\mu\nu}$, $C_{P}^{\mu}$, $B_{P\mu}$)
are recursively determined by the equations of motion. Their single-particle instance corresponds to the respective
polarizations. For example, $H_{p\mu\nu}=h_{p\mu\nu}$ describes a graviton.
In our construction, single-particle states can also be ghosts.

Using \eqref{eq:eom-g}, the recursion for $I_{P}^{\mu \nu}$ is shown to be
\begin{multline}
s_{P} \mathbb{P}_{\mu\nu\rho\sigma} I_{P}^{\rho\sigma}=\frac{1}{2}\sum_{P=Q\cup R}I_{Q}^{\rho\sigma}I_{R}^{\gamma\lambda}\mathcal{V}^{(3)}_{\mu\nu\rho\sigma\gamma\lambda}(Q,R)\\
+\sum_{P=Q\cup R\cup S}I_{Q}^{\rho\sigma}I_{R}^{\gamma\lambda}H_{S}^{\tau\delta}V^{(4)}_{\mu\nu\rho\sigma\gamma\lambda\tau\delta}(Q,R,S)\\
+\sum_{P=Q\cup R\cup S\cup T}I_{Q}^{\rho\sigma}I_{R}^{\gamma\lambda}H_{S}^{\tau\delta}H_{T}^{\alpha\beta}V^{(5)}_{\mu\nu\rho\sigma\gamma\lambda\tau\delta\alpha\beta}(Q,R,S)\\
+\sum_{P=Q\cup R}B_{Q\rho}C_{R}^{\sigma}\mathcal{V}^{(\textrm{g})\rho}_{\mu\nu\sigma}(P,Q), \label{eq:graviton-recursion}
\end{multline}
where we have $s_P=k^2_P$, and
\begin{equation}
\mathbb{P}_{\mu \nu \rho \sigma}=\frac{1}{2}\eta_{\mu\rho}\eta_{\nu\sigma}+\frac{1}{2}\eta_{\mu\sigma}\eta_{\nu\rho}+\frac{1}{(2-d)}\eta_{\mu\nu}\eta_{\rho\sigma}.
\end{equation}
We define also the inverse of $\mathbb{P}_{\mu\nu\rho\sigma}$,
\begin{equation}
(\mathbb{P}^{-1})^{\mu \nu \rho \sigma}=\frac{1}{2}(\eta^{\mu\rho}\eta^{\nu\sigma}+\eta^{\mu\sigma}\eta^{\nu\rho}-\eta^{\mu\nu}\eta^{\rho\sigma}).
\end{equation}
The recursion sums in \eqref{eq:graviton-recursion} go over the possible disjoint, ordered sub-words $\{Q,R,\ldots \}$ that form the word $P=Q\cup R \cup \ldots$. The current $H_{P\mu\nu}$ is implicitly determined by $\mathfrak{g}^{\mu \rho}\mathfrak{g}_{\nu\rho}=\delta^{\mu}_{\nu}$ in terms of $I^{\mu\nu}_P$,
\begin{equation}
H_{P\mu\nu}=\eta_{\mu\rho}I_{P}^{\rho\sigma}\eta_{\sigma\nu}+\eta_{\mu\rho}\sum_{P=Q\cup R}I_{Q}^{\rho\sigma}H_{R\sigma\nu}, \label{eq:inversion}
\end{equation}and the remaining tensor structures in \eqref{eq:graviton-recursion}  are
\begin{multline}
\mathcal{V}^{(3)}_{\mu\nu\rho\sigma\gamma\lambda}(Q,R)= k_{QR\gamma}k_{R\nu}\eta_{\mu \rho}\eta_{\sigma\lambda} - k_{Q\gamma}k_{R\sigma}\eta_{\mu\rho} \eta_{\nu\lambda} \\ +k_{QR\rho}k_{Q\nu}\eta_{\mu \gamma}\eta_{\sigma\lambda}+ \frac{1}{2} \Big( k_{QR\gamma}k_{Q\lambda}\mathbb{P}_{\mu\nu\rho\sigma}
 \\ \vphantom{ \frac{1}{2}}- k_{Q\mu}k_{R\nu}\mathbb{P}_{\rho\sigma\gamma\lambda} + k_{QR\rho}k_{R\sigma}\mathbb{P}_{\mu\nu\gamma\lambda}  - s_{Q}\eta_{\mu \gamma}\mathbb{P}_{\rho\sigma\nu\lambda}  \\- s_{R}\eta_{\mu \rho}\mathbb{P}_{\gamma\lambda\nu\sigma} - s_{QR}\eta_{\sigma\lambda}\mathbb{P}_{\mu\nu\rho\gamma}\Big)+(\mu\leftrightarrow\nu),\label{eq:3pt-vertex}
\end{multline}
\begin{widetext} \noindent
\begin{multline}
V^{(4)}_{\mu\nu\rho\sigma\gamma\lambda\tau\delta}(Q,R,S) = k_{QRS\rho}k_{Q\nu}\eta_{\mu\lambda}\eta_{\gamma\tau}\eta_{\sigma\delta}-k_{Q\lambda}k_{R\sigma}\eta_{\mu\rho}\eta_{\nu\delta}\eta_{\gamma\tau}-\frac{1}{4}k_{Q\mu}k_{RS\nu}\eta_{\rho\lambda}\eta_{\gamma\tau}\eta_{\sigma\delta}\\
 +\frac{1}{2}(k_{QRS\rho}k_{RS\sigma}-k_{R\rho}k_{S\sigma})\eta_{\mu\lambda}\eta_{\nu\delta}\eta_{\gamma\tau}+\frac{1}{2}[(k_{Q}\cdot k_{R})-s_{QRS}]\eta_{\mu\rho}\eta_{\nu\delta}\eta_{\sigma\lambda}\eta_{\gamma\tau}+\frac{1}{2}(k_{Q}\cdot k_{RS})\eta_{\mu\rho}\eta_{\nu\lambda}\eta_{\sigma\delta}\eta_{\gamma\tau}\\
 +\frac{1}{2}\frac{1}{(2-d)}[k_{QRS\rho}k_{R\sigma}\eta_{\mu\nu}\eta_{\gamma\tau}\eta_{\lambda\delta}-k_{Q\mu}k_{R\nu}\eta_{\rho\sigma}\eta_{\gamma\tau}\eta_{\lambda\delta}-(k_{QR}\cdot k_{Q})\eta_{\mu\tau}\eta_{\nu\delta}\eta_{\rho\gamma}\eta_{\sigma\lambda}]\\
 +\frac{1}{2}\frac{1}{(2-d)}[k_{Q\gamma}k_{QR\lambda}\eta_{\mu\tau}\eta_{\nu\delta}\eta_{\rho\sigma}-(k_{QRS}\cdot k_{Q})\eta_{\mu\nu}\eta_{\rho\lambda}\eta_{\gamma\tau}\eta_{\sigma\delta}-s_{Q}\eta_{\mu\lambda}\eta_{\nu\delta}\eta_{\rho\sigma}\eta_{\gamma\tau}]+(\mu\leftrightarrow\nu),
\end{multline}
\begin{multline}
V^{(5)}_{\mu\nu\rho\sigma\gamma\lambda\tau\delta\alpha\beta}(Q,R,S)=\frac{1}{2}\eta_{\mu\alpha}\eta_{\rho\beta}\eta_{\sigma\delta}[(k_{Q}\cdot k_{RS})\eta_{\nu\gamma}\eta_{\lambda\tau}-k_{Q\gamma}k_{S\lambda}\eta_{\nu\tau}]-\frac{1}{2}k_{Q\gamma}k_{R\sigma}\eta_{\mu\tau}\eta_{\nu\alpha}\eta_{\rho\delta}\eta_{\lambda\beta}\\
+\frac{1}{2}\frac{1}{(2-d)}\eta_{\mu\alpha}\eta_{\nu\beta}\eta_{\gamma\tau}[k_{QRS\rho}k_{R\sigma}\eta_{\lambda\delta}-(k_{QRS}\cdot k_{Q})\eta_{\rho\lambda}\eta_{\sigma\delta}]-\frac{1}{4}\frac{1}{(2-d)}k_{Q\mu}k_{R\nu}\eta_{\rho\tau}\eta_{\sigma\delta}\eta_{\gamma\alpha}\eta_{\lambda\beta}+(\mu\leftrightarrow\nu),
\end{multline}
\end{widetext}
\begin{multline}
\mathcal{V}^{(\textrm{g})\rho}_{\mu\nu\sigma}(P,Q) =\frac{1}{2} (k_{Q \mu}-k_{P \mu}) k_{Q \nu} \delta^{\rho}_{\sigma} \\ -\frac{1}{2} k_{P \mu}k_{Q\sigma} \delta^{\rho}_{\nu} +(\mu \leftrightarrow\nu). \label{eq:ghost-vertex}
\end{multline}

\begin{figure}[h]
\def\svgwidth{\linewidth}
\centering{\resizebox{88mm}{!}{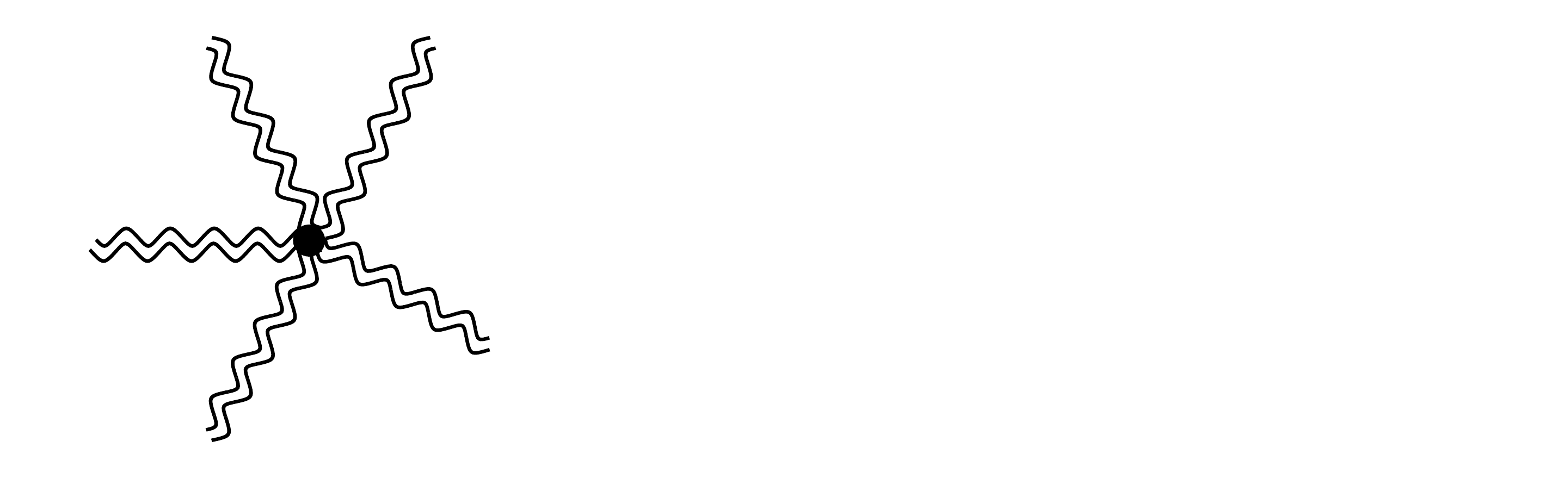}}
\caption{$(n+1)$-graviton vertex.}
\label{fig:Nptvertex}
\end{figure}

These objects are clearly connected to the Feynman vertices of the theory. Using \eqref{eq:inversion}, we can make this more explicit. The recursion \eqref{eq:graviton-recursion} is suggestively rewritten as
\begin{multline}
s_{P} \mathbb{P}_{\mu\nu\rho\sigma} I_{P}^{\rho\sigma} = \sum_{n=2}^{\infty}\sum_{P=P_{1}\cup\ldots\cup P_{n}}\prod_{i=1}^{n}I_{P_{i}}^{\mu_{i}\nu_{i}}\\\times \frac{1}{n!} \mathcal{V}_{\mu \nu \mu_{1}\nu_{1}\cdots\mu_{n}\nu_{n}}^{(n+1)}(P_1,\ldots,P_{n})+\textrm{ghosts}, \label{eq:graviton-perturbiner}
\end{multline}
where $\mathcal{V}^{(3)}$ is given in \eqref{eq:3pt-vertex} and the general expression for $n\geq 3$ is given by
\begin{multline}
\mathcal{V}^{(n+1)}_{\mu\nu\mu_{1}\nu_{1}\cdots\mu_{n}\nu_{n}}(P_1,\ldots,P_{n}) \\=  V^{(4)}_{\mu\nu\mu_{1}\nu_{1}\mu_{2}\nu_{2}\mu_{3}\nu_{n}}(P_1,P_2,P_{3}\cup \ldots \cup P_{n})\prod_{i=3}^{n-1}\eta_{\nu_{i}\mu_{i+1}}\\
+\sum_{k=4}^{n}V^{(5)}_{\mu\nu\mu_{1}\nu_{1}\mu_{2}\nu_{2}\mu_{3}\nu_{k-1}\mu_{k}\nu_{n}}(P_1,P_2,P_{3}\cup \ldots \cup P_{k-1})  \\\times \prod_{i=4}^{k-1} \eta_{\nu_{i-1}\mu_{i}} \prod_{j=k}^{n-1} \eta_{\nu_{j}\mu_{j+1}}+S(1,\ldots,n).\label{eq:npoint-vertex}
\end{multline}
Here, $S(1,\ldots,n)$ denotes the permutations of the labels $\{1,2,\ldots,n\}$. $\mathcal{V}^{(n)}$ corresponds to the $n$-graviton vertex after symmetrizing in $\mu_i \leftrightarrow \nu_i$ (see figure \ref{fig:Nptvertex}). The three-, four-, and five-point vertices are in agreement with the literature \cite{Capper:1973pv,Brandt:1992dk}. 

What makes the multiparticle currents special is the fact that they are related to scattering trees with off-shell external legs  \cite{Gomez:2022dzk}. The amputated correlators are defined through \begin{equation}
\mathcal{M}_N^\textrm{tree}=h^{\mu \nu}_{1} ( s_{2\ldots N} \mathbb{P}_{\mu\nu\rho\sigma} I_{2\ldots N}^{\rho\sigma}).
\end{equation}
Momentum conservation is left implicit, and external legs are off the mass-shell. Tree level amplitudes are then simply obtained after imposing the physical state conditions, i.e. $k^2_p=0$, with $h^{\mu \nu}_p$ traceless and transversal,
\begin{equation}
M_N^\textrm{tree}= \left. \mathcal{M}_N^\textrm{tree} \right|_\textrm{on-shell}.
\end{equation}
We have tested this formula up to seven-point tree level amplitudes, matching numerical computations using the CHY formalism \cite{Cachazo:2013iea}.

\section{Loop recursions}

The off-shell scattering trees  generated by \eqref{eq:graviton-perturbiner} are the intermediate step in the construction of loop integrands. Next, we introduce the loop-closing leg $\ell$ via the identification $I^{\mu \nu}_{\ell P}= h^{\alpha \beta}_{\ell} K^{\mu \nu}_{P \alpha \beta}$, with $K^{\mu \nu}_{P \alpha \beta}=K^{\mu \nu}_{P \alpha \beta}(\ell)$. We then have
\begin{multline}
s_{\ell P} \mathbb{P}_{\mu\nu\rho\sigma} K^{\rho \sigma}_{P \alpha \beta} =I_{P}^{\rho \sigma} \mathcal{V}_{\mu \nu \rho \sigma \alpha \beta }^{(3)}(P,\ell ) \\+\sum_{n=2}^{\infty}\frac{1}{(n-1)!}\sum_{P=P_{1}\cup\ldots\cup P_{n}}\prod_{i=1}^{n-1}I_{P_{i}}^{\mu_{i}\nu_{i}}\\\times \bigg\{\frac{1}{n} I_{P_{n}}^{\mu_{n}\nu_{n}} \mathcal{V}_{\mu \nu \mu_{1}\nu_{1} \cdots\mu_{n}\nu_{n}\alpha \beta }^{(n+2)}(P_1,\ldots,P_{n},\ell ) \\ +K^{\mu_{n}\nu_{n}}_{P_n \alpha \beta} \mathcal{V}_{\mu \nu \mu_{1}\nu_{1}\cdots\mu_{n}\nu_{n}}^{(n+1)}(P_1,\ldots,\ell P_{n})\bigg\}.\label{eq:pre-integrand}
\end{multline}
It directly follows from \eqref{eq:graviton-perturbiner}, after removing $h_{\ell}^{\alpha \beta}$ from both sides and leaving implicit the symmetry in $\alpha \leftrightarrow \beta$. Note there are no ghost external legs yet.

A naive sewing of the two legs $(\mu \nu)$ and $(\alpha \beta)$ would lead to several one-loop integrands. This is achieved by multiplying both sides of  \eqref{eq:pre-integrand} by  the graviton propagator $D^{\mu \nu \alpha \beta}_{\ell} =(\mathbb{P}^{-1})^{\mu \nu \alpha \beta}/s_\ell$. Though we are guaranteed to generate all one-loop diagrams, this procedure does not immediately yield the one-loop correlators upon integration of the loop momentum $\ell^\mu \equiv k^\mu_\ell$. We are clearly overlooking symmetry factors and equivalent-diagram contributions. A natural proposal to try to correct this shortcoming is to define a modified $K_P$, namely
\begin{multline}
\tilde{K}^{\rho \sigma}_{P \alpha \beta}(\ell) =D^{\mu \nu \rho \sigma}_{\ell P}\Big\{ f_1(|P|)I_{P}^{\gamma \lambda} \mathcal{V}_{\mu \nu \gamma \lambda \alpha \beta }^{(3)}(P,\ell ) \\+\sum_{n=2}^{\infty}\frac{1}{(n-1)!}\sum_{P=P_{1}\cup\ldots\cup P_{n}}\prod_{i=1}^{n-1}I_{P_{i}}^{\mu_{i}\nu_{i}}\\\times \Big[\frac{1}{n} f_{n} I_{P_{n}}^{\mu_{n}\nu_{n}} \mathcal{V}_{\mu \nu \mu_{1}\nu_{1} \cdots\mu_{n}\nu_{n}\alpha \beta }^{(n+2)}(P_1,\ldots,P_{n},\ell ) \\ \vphantom{\frac{1}{n}}+g_{n} \tilde{K}^{\mu_{n}\nu_{n} }_{P_n \alpha \beta}(\ell) \mathcal{V}_{\mu \nu \mu_{1}\nu_{1}\cdots\mu_{n}\nu_{n}}^{(n+1)}(P_1,\ldots,\ell P_{n})\Big] \Big\}.
\label{eq:modified-integrand}
\end{multline}
The coefficients $f_{n}=f_{n}(|P_j|)$ and $g_{n}=g_{n}(|P_j|)$ depend on the subword lengths $\{|P_1|,\ldots,|P_n|\}$, and should balance out any overcounting ($0\leq f_{n},g_{n}\leq 1$). The graviton-loop integrand contribution to the one-loop amplitude would then be given by
\begin{equation}
\mathcal{I}^{\textrm{graviton}}_N(\ell) = \tilde{K}^{\mu \nu}_{1\ldots N \mu \nu}(\ell), \label{eq:1loop-graviton}
\end{equation}
with conserved momentum $k_{1\ldots N}=0$. 

The fact that there exist solutions to $f_n$ and $g_n$ is noteworthy. Yet more remarkably, they take a simple form:
\begin{eqnarray}
f_n = \frac{1}{2}, &\hspace{1cm}& g_n= \frac{|P_n|}{|P|}.
\end{eqnarray}
In \eqref{eq:modified-integrand}, the coefficients $f_n$ accompany tadpole diagrams, in which the loop is formed via the sewing of two legs of the same vertex. The symmetry factor of such diagrams is simply $1/2$. It is the same factor of the so-called bubble diagrams, in which the loop contains only two vertices. The solution for $g_n$ takes care of the repeated diagrams that would otherwise be generated when using \eqref{eq:pre-integrand}.

Now, in order to compute the ghost loop, we need first the related tree level recursions. Their derivation from the respective equations of motion is straightforward. For the $c$ ghost, for instance, we obtain
\begin{equation}
s_{P}C^{\rho}_{P} = \sum_{P=Q\cup R}I_{Q}^{\mu \nu} C_{R}^{\sigma} \mathcal{V}^{(\textrm{g})\rho}_{\mu\nu\sigma}(Q,P), \label{eq:ghost-perturbiner}
\end{equation}
The field theory vertex $\mathcal{V}^{(\textrm{g})\rho}_{\mu\nu\sigma}$ is defined in \eqref{eq:ghost-vertex}, and there is a similar construction for $B_{P}$.

We then introduce the loop-closing leg $\ell$ via the identification $C^{\mu}_{\ell P}=  J^{\mu}_{P\nu}  c^{\nu}_{\ell}$, with $J^{\mu}_{P\nu}=J^{\mu}_{P\nu} (\ell)$. From \eqref{eq:ghost-perturbiner} we obtain
\begin{multline}
s_{\ell P} J^{\rho}_{P \sigma} =
I_{P}^{\mu \nu}\mathcal{V}^{(\textrm{g})\rho}_{\mu\nu\sigma}(P,\ell P)
\\+ \sum_{P=Q\cup R}I_{Q}^{\mu \nu}J^{\gamma}_{R \sigma} \mathcal{V}^{(\textrm{g})\rho}_{\mu\nu\gamma}(Q,\ell P).
\end{multline}

We can mimic the graviton loop analysis and sew the ghost and antighost legs with the ghost propagator $\delta^{\sigma}_{\rho}/ \ell^2$. This would again lead to an overcounting of different contributions, so we introduce a modified current
\begin{multline}
s_{\ell P} \tilde{J}^{\rho}_{P \sigma} =
I_{P}^{\mu \nu}\mathcal{V}^{(\textrm{g})\rho}_{\mu\nu\sigma}(P,\ell P)
\\+ \sum_{P=Q\cup R} \frac{|Q|}{|P|} I_{Q}^{\mu \nu}\tilde{J}^{\gamma}_{R \sigma} \mathcal{V}^{(\textrm{g})\rho}_{\mu\nu\gamma}(Q,\ell P),
\label{eq:ghost-loop-current}
\end{multline}
where we found that the factor $|Q|/|P|$ is needed to balance the repeated diagrams. After imposing momentum conservation $k_P=0$ on the external legs, we can finally define the ghost-loop integrand:
\begin{equation}
\mathcal{I}^{\textrm{ghost}}_N(\ell) = \tilde{J}^{\mu}_{1 \ldots N\mu}(\ell).\label{eq:1loop-ghost}
\end{equation}

The complete one-loop integrand for the $N$-graviton amputated correlator $\mathcal{M}^{1\textrm{-loop}}_N$ is then
\begin{eqnarray}
\mathcal{I}_N(\ell) &=& \mathcal{I}^{\textrm{graviton}}_{N}(\ell)-\mathcal{I}^{\textrm{ghost}}_{N}(\ell), \label{eq:loop-integrand}\\
\mathcal{M}^{1\textrm{-loop}}_N &=& \int \frac{d^d\ell}{(2\pi)^d}\, \mathcal{I}_N(\ell). \label{eq:1loop-correlator}
\end{eqnarray}
As usual, the ghost loop comes with a negative sign.

If we are interested in computing one-loop amplitudes, we can just impose the on-shell condition on external legs. In particular, our prescription enables a clean removal of tadpoles diagrams and external-leg bubbles. The amplitude is defined as
\begin{multline}
M^{1\textrm{-loop}}_N = \frac{1}{N} \int \frac{d^d\ell}{(2\pi)^d}  \\
\times \bigg\{ \sum_{n=2}^{N}\frac{1}{(n-1)!}\sum_{\substack{1\ldots N=P_{1}\cup\ldots\cup P_{n}\\|P_n|>1}} \prod_{i=1}^{n-1}I_{P_{i}}^{\mu_{i}\nu_{i}} \\
\times |P_n| \tilde{K}^{\mu_{n}\nu_{n} }_{P_n \alpha \beta} (\mathbb{P}^{-1})^{\alpha \beta \mu \nu} \mathcal{V}_{\mu \nu \mu_{1}\nu_{1}\cdots\mu_{n}\nu_{n}}^{(n+1)}(P_1,\ldots,\ell P_{n})
\\- \sum_{\substack{1\ldots N=Q\cup R \\|R|>1}} |Q| I_{Q}^{\mu \nu}\tilde{J}^{\sigma}_{R \rho} \mathcal{V}^{(\textrm{g})\rho}_{\mu\nu\sigma}(Q,\ell P)\bigg\}. \label{eq:Npoint-amplitude}
\end{multline}
This prescription follows directly from \eqref{eq:1loop-correlator}, with two modifications in equations \eqref{eq:1loop-graviton} and \eqref{eq:1loop-ghost}. First, we delete the terms involving vertex self-contractions. These correspond, respectively, to the terms multiplied by $f_n$ in \eqref{eq:modified-integrand} and the first line on the right-hand-side of \eqref{eq:ghost-loop-current}. Second, we remove double contractions between cubic vertices with one external leg and any other vertex, which trace back to $|P_n|=|R|=1$.

More generally, it is straightforward to identify specific diagrams in \eqref{eq:loop-integrand} through the propagator structure of the different terms.  Their extraction can be easily automated. In what follows, we will present some examples to illustrate our proposal.

\section{Some examples}

The simplest integrand we could compute using \eqref{eq:loop-integrand} is the case $N=1$, which yields one-graviton tadpoles,
\begin{multline}
\mathcal{I}_1(\ell)= \frac{1}{2 \ell^2} h_{1}^{\mu \nu} (\mathbb{P}^{-1})^{\rho \sigma \alpha \beta} \mathcal{V}_{\rho \sigma \mu \nu  \alpha \beta }^{(3)}(1,\ell ) \\ -\frac{1}{\ell^2} h_{1}^{\mu \nu} \mathcal{V}^{(\textrm{g})\rho}_{\mu\nu\rho}(1,\ell 1),
\end{multline}
with $k_1=0$.

The more interesting case would be the $2$-graviton correlator, with integrand
\begin{equation}
\mathcal{I}_2(\ell)=\mathcal{I}^{\textrm{tp}}_2+\mathcal{I}^{\textrm{se}}_2. \label{eq:2gra-integrand}
\end{equation}
The first term consists of two-graviton tadpoles,
\begin{multline}
\mathcal{I}^{\textrm{tp}}_2 = \frac{1}{2\ell^2}  (\mathbb{P}^{-1})^{\rho \sigma \alpha \beta} \big\{I_{12}^{\mu \nu}\mathcal{V}_{\rho \sigma \mu \nu  \alpha \beta }^{(3)}(12,\ell )\\ +  h_{1}^{\mu_{1}\nu_{1}} h_{2}^{\mu_{2}\nu_{2}} \mathcal{V}_{\rho \sigma \mu_{1}\nu_{1} \mu_{2}\nu_{2}\alpha \beta }^{(4)}(1,2,\ell)\big\} \\ -\frac{1}{\ell^2}I_{12}^{\mu \nu}\mathcal{V}^{(\textrm{g})\rho}_{\mu\nu\rho}(12,\ell 12) .
\end{multline}
The second term in \eqref{eq:2gra-integrand} corresponds to the graviton self-energy, one contribution from the graviton loop and one from the ghost loop,
\begin{multline}
\mathcal{I}^{\textrm{se}}_2 =\frac{h_{1}^{\mu_1 \nu_1} h_{2}^{\mu_{2}\nu_{2}}}{ \ell^2 (\ell +k_1)^2}     \Big\{ \frac{1}{4} (\mathbb{P}^{-1})^{\mu \nu \gamma \lambda} (\mathbb{P}^{-1})^{\alpha \beta \rho \sigma } \\
\times \mathcal{V}_{\mu \nu \mu_1 \nu_1 \alpha \beta }^{(3)}(1,\ell )   \mathcal{V}_{\rho\sigma \mu_{2}\nu_{2}\gamma \lambda}^{(3)}(2,\ell 1) \\ - \frac{1}{2} \mathcal{V}^{(\textrm{g})\sigma}_{\mu_1\nu_1\rho}(1,\ell 1) \mathcal{V}^{(\textrm{g})\rho}_{\mu_2\nu_2\sigma}(2,\ell 12) \Big\}+(1\leftrightarrow 2). \label{eq:2pt-1loop-integrand}
\end{multline}
The symmetry in the exchange of particles $1$ and $2$ is explicit. The exchanged integrands are related by a redefinition of the loop momentum. The graviton self-energy is computed to be
\begin{equation}
\mathcal{M}^{1\textrm{-loop}}_2 = h_{1}^{\mu \nu} h_{2}^{\rho \sigma}  \Pi_{\mu \nu \rho \sigma}.
\end{equation}
The expression for $\Pi_{\mu \nu \rho \sigma}$ is more commonly presented in terms of the propagator correction, $Q^{\mu \nu \rho \sigma} =  D^{\mu \nu \alpha \beta}_1 D^{\rho \sigma \gamma \lambda}_2 \Pi_{\alpha \beta \gamma \lambda}$, given by
\begin{multline}
Q^{\mu \nu \rho \sigma} = B_0\big\{c_{1}\Pi^{\mu\nu}\Pi^{\alpha\beta}  -\Pi^{\mu\nu}k^{\alpha}k^{\beta}-\Pi^{\alpha\beta}k^{\mu}k^{\nu} \\ - c_{2}(\Pi^{\mu\alpha}\Pi^{\nu\beta}+\Pi^{\mu\beta}\Pi^{\nu\alpha}) \big\}, \label{eq:prop-correction}
\end{multline}
with $k_1 = -k_2 = k$, $\Pi_{\mu \nu}= k^2 \eta_{\mu \nu} - k_\mu k_\nu$,  and
\begin{eqnarray}
c_{1} & = & \frac{[64+(94+(20+(d-13)d)d)d]}{8(d^2-1)},\\
c_{2} & = & \frac{[16 + (15- (9+4d)d)d]}{8(d^2-1)}.
\end{eqnarray}
The coefficient $B_0=B_0(k^2)$ is expressed in terms of the scalar integral
\begin{equation}
B_0 =  \frac{1}{2 k^4}\int \frac{d^d \ell}{(2 \pi)^d} \frac{1}{\ell^2(\ell + k)^2},
\end{equation}
which usually appears when using dimensional regularization (\cite{Capper:1974dc} and references therein). In this case, the tadpole contributions are regularized to zero.

Equation \eqref{eq:prop-correction} can be validated through the Slavnov-Taylor identity involving the divergence of the two-graviton correlator. It is easy to show that
\begin{equation}
k_\mu Q^{\mu \nu \rho \sigma} = - B_0 k^2  k^\nu \Pi^{\rho \sigma},
\end{equation}
The complete identity can be found in \cite{Capper:1974vb}, which we have verified using our proposal. The weaker identity  $k_\mu k_{\rho}Q^{\mu \nu \rho \sigma}=0$ trivially follows \cite{Capper:1973pv}.

Though the three-graviton one-loop amplitude is zero, we have checked that the off-shell integrands in our proposal are qualitatively and quantitavely correct. We have also obtained the four-graviton amplitude and it matches the literature \cite{Dunbar:1994bn}. Perhaps the easiest way to see this is through the integrand. Equation \eqref{eq:npoint-vertex} provides the correct Feynman vertices of the theory, while equation \eqref{eq:Npoint-amplitude} generates the complete diagrammatic expansion.

Finally, we have shown using symbolic computation that \eqref{eq:loop-integrand} leads to one-loop integrands with the correct symmetry factors. This includes all diagrams with external gravitons up to five-points, and randomly selected diagrams in six-, seven-, and eight-points. We are making some of these computations available through a sample Mathematica notebook coupled to the \href{https://arxiv.org/}{arXiv} submission.

\section{Final remarks}

In this work we have proposed an algebraic recursion to generate the complete one-loop integrands of $N$-graviton correlators. They can be extracted from equation \eqref{eq:loop-integrand} by replacing the external polarizations $h^{\mu \nu}_p$ by the respective propagators $D^{\mu \nu \rho \sigma}_{p}$. This recursion is an off-shell evolution of the perturbiner method. As in the tree level case, the combinatoric burden that comes with the traditional Feynman graphs approach is seamlessly overcome. In addition, the $N$-graviton field theory vertices manifestly appear in the recursion\footnote{All multiplicity graviton/ghost vertices have been written in the past (see e.g. \cite{Latosh:2023zsi}). Up to our knowledge this is the first time they are explicitly presented using the metric density, which take a radically simpler form. See also \cite{Cho:2022faq} for the weak-field expansion of the related Lagrangian.}. The analysis of divergencies from the loop momentum integral is identical to the one using Feynman diagrams. And it is interesting to point out that dimensional regularization can be directly applied to our output.

Our proposal outperforms the previous results of \cite{Gomez:2022dzk} in two important points. First, it preserves the simple structure of the gravitational perturbiner \cite{Gomez:2021shh}, without any other decoration in the sums over subwords. Second, it can be applied to theories with mixed or no color structure at all, in particular with arbitrarily high number of vertices.

There are a couple of questions of immediate interest to be addressed. The extension of our formulas to higher loops would be an impressive achievement. The strategy would be similar: (i) singling out loop-closing legs, (ii) sewing them, and (iii) introducing modified recursions to deal with diagram overcounting and symmetry factors. Besides uncovering a more structured account of the diagram combinatorics, one would be in a good position to identify multi-loop patterns of the multiparticle currents. Relatedly, the application of the method to more diverse field theories would help to better explain the combinatoric factors within the algebraic recursion. We already know that the perturbiner naturally describes gravity coupled to matter \cite{Gomez:2021shh}. Thus we are on a good path to analyze matter loops as well, which are more interesting in the gravitational waves context. We hope to report some progress in these directions soon.

As a computational tool, the loop pertubiner will also find applications in gravitational EFTs. The program started by Donoghue \cite{Donoghue:1993eb,Donoghue:1994dn} offers a more pragmatic view on quantum effects in gravity \cite{Bjerrum-Bohr:2002fji,Bjerrum-Bohr:2002gqz}. In this direction, we also highlight the mixing of classical and quantum effects at loop level \cite{Holstein:2004dn}. Perhaps less obviously, our results might be used in the recent efforts to model black-hole binary mergers using scattering amplitudes, though they are mostly based on on-shell methods (e.g. \cite{Bjerrum-Bohr:2013bxa}, see also \cite{Bjerrum-Bohr:2022blt} for a review).

\begin{acknowledgments}
We would like to thank N. Emil J. Bjerrum-Bohr, P. H. Damgaard, J. Donoghue, and C. Schubert for useful feedback on the manuscript. The work of HG and RLJ was partially supported by the European Structural and Investment Funds and the Czech Ministry of Education, Youth and Sports (project FORTE CZ.02.01.01/00/22\_008/0004632). CLA was partially supported during the latest stages of this work by the Munich Institute for Astro-, Particle and BioPhysics (MIAPbP) which is funded by the Deutsche Forschungsgemeinschaft (DFG, German Research Foundation) under Germany's Excellence Strategy - EXC-2094 - 390783311. 
\end{acknowledgments}

\pagebreak

\end{document}

%% file: Npt-vertex.pdf_tex
\begingroup%
  \makeatletter%
  \providecommand\color[2][]{%
    \errmessage{(Inkscape) Color is used for the text in Inkscape, but the package 'color.sty' is not loaded}%
    \renewcommand\color[2][]{}%
  }%
  \providecommand\transparent[1]{%
    \errmessage{(Inkscape) Transparency is used (non-zero) for the text in Inkscape, but the package 'transparent.sty' is not loaded}%
    \renewcommand\transparent[1]{}%
  }%
  \providecommand\rotatebox[2]{#2}%
  \newcommand*\fsize{\dimexpr\f@size pt\relax}%
  \newcommand*\lineheight[1]{\fontsize{\fsize}{#1\fsize}\selectfont}%
  \ifx\svgwidth\undefined%
    \setlength{\unitlength}{1382.69182452bp}%
    \ifx\svgscale\undefined%
      \relax%
    \else%
      \setlength{\unitlength}{\unitlength * \real{\svgscale}}%
    \fi%
  \else%
    \setlength{\unitlength}{\svgwidth}%
  \fi%
  \global\let\svgwidth\undefined%
  \global\let\svgscale\undefined%
  \makeatother%
  \begin{picture}(1,0.3170428)%
    \lineheight{1}%
    \setlength\tabcolsep{0pt}%
    \put(0,0){\includegraphics[width=\unitlength,page=1]{Npt-vertex.pdf}}%
    \put(-0.00120067,0.15095088){\makebox(0,0)[lt]{\lineheight{1.25}\smash{\begin{tabular}[t]{l}$\mu \nu$\end{tabular}}}}%
    \put(0.08276333,0.00341132){\makebox(0,0)[lt]{\lineheight{1.25}\smash{\begin{tabular}[t]{l}$\mu_n \nu_n$\end{tabular}}}}%
    \put(0.07891998,0.30531086){\makebox(0,0)[lt]{\lineheight{1.25}\smash{\begin{tabular}[t]{l}$\mu_1 \nu_1$\end{tabular}}}}%
    \put(0.23583668,0.30605314){\makebox(0,0)[lt]{\lineheight{1.25}\smash{\begin{tabular}[t]{l}$\mu_2 \nu_2$\end{tabular}}}}%
    \put(0.27822159,0.05895625){\makebox(0,0)[lt]{\lineheight{1.25}\smash{\begin{tabular}[t]{l}$\mu_i \nu_i$\end{tabular}}}}%
    \put(0.35546592,0.17095511){\makebox(0,0)[lt]{\lineheight{1.25}\smash{\begin{tabular}[t]{l}$=\mathcal{V}^{(n+1)}_{\mu \nu (\mu_1 \nu_1) \ldots (\mu_n \nu_n)}(P_1,\ldots,P_n).$\end{tabular}}}}%
    \put(0,0){\includegraphics[width=\unitlength,page=2]{Npt-vertex.pdf}}%
  \end{picture}%
\endgroup%